\documentclass[twocolumn,superscriptaddress,amsmath,amssymb,longbibliography,prd,10pt,aps]{revtex4-2}

\newcommand{\Eb}{E$_{\text{blip}}$\xspace}

\usepackage{xcolor,color}
\usepackage{graphicx}
\usepackage{url}
\usepackage{mathtools}
\usepackage{makecell}
\usepackage{dcolumn}
\usepackage{bm}
\usepackage{xspace}
\usepackage{multirow}

\usepackage[colorlinks = true,
            linkcolor = red,
            urlcolor  = blue,
            citecolor = red,
            anchorcolor = blue]{hyperref}

\usepackage{lineno}

\graphicspath{{figures}}

\begin{document}

\title{Measurements of Pion and Muon Nuclear Capture at Rest on Argon in the LArIAT Experiment}

\newcommand{\UFederalABC}{Universidade Federal do ABC, Santo Andr\'{e}, SP 09210-580, Brasil}
\newcommand{\UCampinas}{Universidade Estadual de Campinas, Campinas, SP 13083-859, Brasil}
\newcommand{\UChicago}{University of Chicago, Chicago, IL 60637, USA}
\newcommand{\cincinnati}{University of Cincinnati, Cincinnati, OH 45221, USA}
\newcommand{\FNAL}{Fermi National Accelerator Laboratory, Batavia, IL 60510, USA}
\newcommand{\UFG}{Universidade Federal de Goi\'{a}s, Goi\'{a}s, CEP 74690-900, Brasil}
\newcommand{\UTArlington}{University of Texas at Arlington, Arlington, TX 76019, USA} 
\newcommand{\Boston}{Boston University, Boston, MA 02215, USA}
\newcommand{\michigan}{Michigan State University, East Lansing, MI 48824, USA}
\newcommand{\duluth}{University of Minnesota, Duluth, Duluth, MN 55812, USA}
\newcommand{\infn}{Istituto Nazionale di Fisica Nucleare (INFN), Rome 00186, Italy}
\newcommand{\louisiana}{Louisiana State University, Baton Rouge, LA 70803, USA}
\newcommand{\manchester}{University of Manchester, Manchester M13 9PL, UK} 
\newcommand{\Syracuse}{Syracuse University, Syracuse, NY 13244, USA}
\newcommand{\utaustin}{University of Texas at Austin, Austin, TX 78712, USA}
\newcommand{\ucollegelondon}{University College London, London WC1E 6BT, UK}
\newcommand{\williamandmary}{College of William \& Mary, Williamsburg, VA 23187, USA}
\newcommand{\Yale}{Yale University, New Haven, CT 06520, USA}
\newcommand{\kek}{High Energy Accelerator Research Organization (KEK), Tsukuba 305-0801, Japan}
\newcommand{\IIT}{Illinois Institute of Technology, Chicago, IL 60616, USA}
\newcommand{\LANL}{Los Alamos National Laboratory, Los Alamos, NM 87545, USA}
\newcommand{\Manchester}{University of Manchester, Manchester M13 9PL, UK}
\newcommand{\Harvard}{Harvard University, Cambridge, MA 02138, USA}
\newcommand{\UCSB}{University of California, Santa Barbara, CA 93106, USA}
\newcommand{\UCDavis}{University of California, Davis, CA 95616, USA}
\newcommand{\Edinburgh}{University of Edinburgh, Edinburgh, EH8 9YL, UK}


\author{M. A. Hernandez-Morquecho}\affiliation{\IIT}
\author{R.~Acciarri} \affiliation{\FNAL}
\author{J.~Asaadi}\affiliation{\UTArlington}
\author{M.~Backfish}\altaffiliation{Current: \UCDavis}\affiliation{\FNAL}
\author{W.~Badgett}\affiliation{\FNAL}
\author{V.~Basque} \affiliation{\FNAL}
\author{F.~d.~M.~Blaszczyk}\affiliation{\FNAL}
\author{W.~Foreman} \altaffiliation{Current: \LANL}\affiliation{\IIT}
\author{R.~Gomes} \affiliation{\UFG}
\author{E.~Gramellini}\affiliation{\Yale}
\author{J.~Ho} \altaffiliation{Current:\Harvard}\affiliation{\UChicago}
\author{E.~Kearns} \affiliation{\Boston}
\author{E.~Kemp} \affiliation{\UCampinas}
\author{T.~Kobilarcik}\affiliation{\FNAL}
\author{M.~King} \affiliation{\UChicago}
\author{B.~R.~Littlejohn} \affiliation{\IIT}
\author{X.~Luo} \affiliation{\UCSB}
\author{A.~Marchionni} \affiliation{\FNAL}
\author{C.~A.~Moura} \affiliation{\UFederalABC}
\author{J.~L.~Raaf} \affiliation{\FNAL}
\author{D.~W.~Schmitz}\affiliation{\UChicago}
\author{M.~Soderberg}\affiliation{\Syracuse}
\author{J.~M.~St.~John}\affiliation{\FNAL}
\author{A.~M.~Szelc}\affiliation{\Edinburgh}
\author{T.~Yang}\affiliation{\FNAL}

\collaboration{LArIAT Collaboration}
\thanks{lariat\_authors@fnal.gov}\noaffiliation

\date{\today}

\begin{abstract}
We report the measurement of the final-state products of negative pion and muon nuclear capture at rest on argon by the LArIAT experiment at the Fermilab Test Beam Facility.  
We measure a population of isolated MeV-scale energy depositions, or blips, in 296 LArIAT events containing tracks from stopping low-momentum pions and muons.  
The average numbers of visible blips are measured to be 0.74 $\pm$ 0.19 and 1.86 $\pm$ 0.17 near muon and pion track endpoints, respectively.  
The 3.6$\sigma$ statistically significant difference in blip content between muons and pions provides the first demonstration of a new method of pion-muon discrimination in neutrino liquid argon time projection chamber experiments.  
LArIAT Monte Carlo simulations predict substantially higher average blip counts for negative muon (1.22 $\pm$ 0.08) and pion (2.34 $\pm$ 0.09) nuclear captures.  
We attribute this difference to Geant4's inaccurate simulation of the nuclear capture process.  
\end{abstract}

\maketitle

Liquid argon time projection chambers (LArTPCs) present a unique combination of large detector mass, excellent spatial resolution, and low detection thresholds that make them attractive for novel particle and nuclear physics measurements.  
By testing single-phase LArTPC capabilities at the MeV and sub-MeV scale -- the lowest end of their dynamic range in energy -- neutrino experiments have provided first measurements of final-state neutrons from neutrino-argon nuclear interactions~\cite{argo_mev}, performed sensitive searches for millicharged particles~\cite{argo_mcp}, and measured radioactive contaminants present in large neutrino LArTPC detectors~\cite{ub_radon,MicroBooNE:2023ftv}.  
Other recent literature has more broadly explored the range of potential applications of low energy thresholds and MeV-scale energy reconstruction in neutrino LArTPC physics~\cite{Castiglioni:2020tsu,DUNE:2020zfm,leplar_paper,Q-Pix:2022zjm,Bezerra:2023gvl}. 

A subset of these studies have highlighted the potential value of low-energy signatures in performing particle and charge-sign identification for pions ($\pi$) and muons ($\mu$) in LArTPCs~\cite{Castiglioni:2020tsu,leplar_paper}.  
As they range out in large detectors, $\pi^-$ and $\mu^-$ will preferentially experience nuclear capture: $\pi^-$\,+\,$^{40}$Ar\,$\rightarrow\,^{40}$Cl$^*$ occurs nearly 100\% of the time for stopped $\pi^-$ in argon while $\mu^-$\,+\,$^{40}$Ar\,$\rightarrow\,^{40}$Cl$^*$\,+\,$\nu_{\mu}$ occurs for 70-75\% of stopped $\mu^-$~\cite{lariat_michels,Castiglioni:2020tsu}.  
Nuclear captures generate MeV-scale deexcitation products ($\gamma$ rays and neutrons), which interact in the surrounding LAr to produce displaced energy depositions reconstructed as ‘blips’ of charge spanning only a few readout wires.  
This distinct final-state topology differs from that of free $\pi^+$/$\mu^+$ decay.
Likewise, since pions deposit more rest-mass energy in capturing nuclei than muons, more blip features should accompany their capture.  
Muon charge-sign selection offers obvious advantages for beam neutrino oscillation physics by reducing wrong-sign contamination during antineutrino-mode run configurations~\cite{NOvA:2019cyt,T2K:2020nqo}.  
Capture-based discrimination may also be useful in new physics or rare process searches involving $\pi/\mu$ final states~\cite{Breitbach:2021gvv,Abdullahi:2022jlv,Altmannshofer:2014pba,deGouvea:2018cfv,MicroBooNE:2019izn, T2K:2019jwa, MicroBooNE:2022ctm}.

Measurements of the products of $\pi/\mu$ nuclear capture at rest on argon are scarce: only one measurement for $\mu$ reports the prevalence of various unstable final state nuclei~\cite{muAr}, and no measurements exist for $\pi$.  
Existing predictions of final states in particle transport simulations such as Geant4~\cite{g4} and \texttt{FLUKA}~\cite{Bohlen:2014buj} are based on nuclear models benchmarked to measurements of various medium-energy nuclear processes on other target nuclei.  
Beyond testing the accuracy of particle transport simulations, a measurement of the products of muon nuclear capture at rest provides insight into the related weak process of low-energy muon neutrino charged-current absorption~\cite{Volpe:2000zn}.  
An argon-based $\pi/\mu$ nuclear capture measurement thus serves as a reference point for low-energy neutrino interaction generators~\cite{Gardiner:2021qfr} used for performing neutrino astrophysics in large LArTPCs~\cite{dune_solar,DUNE:2020zfm,DUNE:2023rtr}.  
It provides a similar reference point for aspects of high-energy neutrino interaction models~\cite{Andreopoulos:2009rq,BUSS20121,GOLAN2012499,Hayato:2021heg}.  

In this Letter, we present measurements of the final-state products of $\pi^-$ and $\mu^-$ nuclear capture at rest ($\pi$CAR and $\mu$CAR) on argon using the Liquid Argon In A Testbeam (LArIAT) experiment at Fermilab.  
Final-state $\gamma$ ray and neutron content is reported in terms of the properties of reconstructed low-energy mm-scale blips generated by these particles in LArTPC events. 
Blip counts are found to be higher in events containing $\pi$CAR and $\mu$CAR than in background events containing only through-going beam particles.  A higher count in $\pi$CAR than in $\mu$CAR events provides the first demonstration of blip-based particle discrimination.  
LArIAT Monte Carlo (MC) simulations predict higher blip counts near nuclear captures than observed in data, suggesting improper modelling of nuclear deexcitation processes in the Geant4 particle transport code.  


The LArIAT experiment was operated from 2015 to 2017 in a tertiary particle beamline in the MC7 hall at Fermilab's Test Beam Facility (FTBF).  
Detailed descriptions of the beam and LArTPC are given in Ref.~\cite{lariat_detpaper}.  
Tertiary beam particles ($e^{\pm}$, $\mu^{\pm}$, $\pi^{\pm}$, $p^{\pm}$, $K^{\pm}$, etc.) were produced for LArIAT by colliding a 64~GeV/c peak momentum $\pi^+$-dominated secondary beam into a copper target.  
Boosted, charged products were collimated and then steered via two electromagnets through beam instrumentation including four position-measuring multi-wire proportional chambers (MWPCs) and two time-of-flight (TOF) scintillator paddles.  
Tuning of beam momentum and charge sign was performed by adjusting the electromagnet current.  
The momentum of individual particles, ranging between 275-1400 MeV/c, was determined by measuring bending radius using two MWPCs upstream and downstream from the electromagnet.  
Particle momenta upon entering the LArIAT LArTPC are lower than beam-reported values due to energy loss 
in materials between the downstream wire chamber and the active TPC volume.  
Combined momentum and TOF measurements enabled clean separation of beam $\pi/\mu/e$ from heavier particle species.  
While particles primarily traverse the beamline individually, pile-up muons and neutrons produced in the primary or secondary targets are also present in MC7 and are not monitored by LArIAT beamline instrumentation.  

The LArIAT cryostat is located approximately 10 cm past the downstream-most TOF detector.  
Filled with 0.76~t of liquid argon, it contains the LArIAT LArTPC, a 47$\times$40$\times$90 cm$^3$, 0.24~t active chamber equipped with a uniform 490 V/cm electric field along the drift direction ($x$ in LArIAT's coordinate system). 
The TPC electric field is perpendicular to the beam particle direction ($z$).  
Ionization generated by beam particles is drifted a maximum $x$ distance of 47~cm towards two planes of conducting sense wires.  
For Run-II, each plane is made of 240 wires spaced 4 mm apart and oriented $\pm$60$^{\circ}$ with respect to the vertical ($y$) direction. 
Drifting charge produces signals on the two instrumented wire planes by passing by the first (induction plane) and collecting on the second (collection plane). 
These signals are amplified, shaped, and digitized to generate 3072-sample waveforms with a sampling period of 0.128~$\mu$s per time tick, for a total readout time of 393~$\mu$s -- sufficient to record all ionization activity occurring within the full 320~$\mu$s drift period as well as some ionization occurring before and after the beam-triggered event.
Triggering of TPC readout is initiated by coincident signals from the FTBF primary beamline, all MWPCs and TOF beamline instruments, and the LArTPC's scintillation light collection system~\cite{lariat_michels}.  
Roughly 20-50~triggers were recorded over each 4.2~second beam spill  delivered to LArIAT once every 60.5 seconds, with 10 triggers free of beam activity (referred to as pedestal triggers) recorded before each spill.  

For this study, we use data from LArIAT's Run II-A and II-B campaigns, collected between February and July 2016.  
Since the study concerns low-momentum beam $\mu^-$ and $\pi^-$ capable of stopping in the TPC and undergoing nuclear capture, data from LArIAT's low-energy negative beam tuning were considered.  
We discarded runs exhibiting electron lifetimes lower than 320~$\mu$s, which is comparable to the cathode-to-anode drift time.  
These requirements leave a total of 62,452 beam and 201,593 pedestal triggers for analysis.  

Procedures for reconstructing high-level physics objects from LArIAT wire ADC waveforms are typical of a single-phase LArTPC~\cite{Baller_2017}, and are described in detail in Ref.~\cite{LArIAT:2021yix}.  
After de-convolving collection and induction plane ADC waveforms using the known response function of LArIAT's electronics, output unipolar signals are scanned for threshold-crossing features, or hits, which are assigned time, width, and amplitude attributes using a Gaussian fit.  
Hit amplitudes can be converted into reconstructed collected charge using electron lifetime and per-channel gain calibration constants derived using TPC-traversing cosmic and beam muons. 
Linearly arranged wire- and time-adjacent hit groups on the two wire planes are matched and clustered to form 3D reconstructed tracks~\cite{Antonello:2012hu}.  

All hits that were not directly included in reconstructed 3D tracks $>$5 cm in length are used as input to reconstruct isolated low energy depositions, or blips. 
Blip reconstruction in LArIAT proceeds similarly to that reported by previous experiments~\cite{argo_mev, MicroBooNE:2023ftv}.  
Time- and wire-adjacent hits are grouped into 2D clusters, which are matched between planes to form a 3D blip using the coincident time ticks and recorded charges.  
Topologies for reconstructed blips vary from compact mm-scale objects with a single hit on each plane to comparatively extended objects with cm-scale blob-like or track-like attributes.  
A blip’s $y$ and $z$ coordinates are determined by the location of crossing wires associated with the blip’s highest-amplitude hit from each plane, while the blip’s $x$ coordinate is calculated based on its drift time and the measured electron drift velocity.  
To reconstruct a blip’s electron-equivalent deposited energy, \Eb, a linear charge-to-energy conversion is used based on the integrated charge on the collection plane~\cite{MicroBooNE:2023ftv}.  
Resulting blip reconstruction efficiency at the dataset's mean electron lifetime of 450 $\mu$s, determined using MC simulations of single uniformly-distributed electrons in the active TPC, exceeds 95\% above 0.3~MeV, while dropping to ~50\% at 0.22~MeV and approaching 0\% at 0.13~MeV.
Blip efficiencies in \Eb and in $x$ vary from run to run due to large lifetime variations over the data-taking period, an effect that is accounted for in LArIAT beam MC simulations and systematic uncertainty estimates.  

\begin{figure}
\includegraphics[width=\columnwidth]{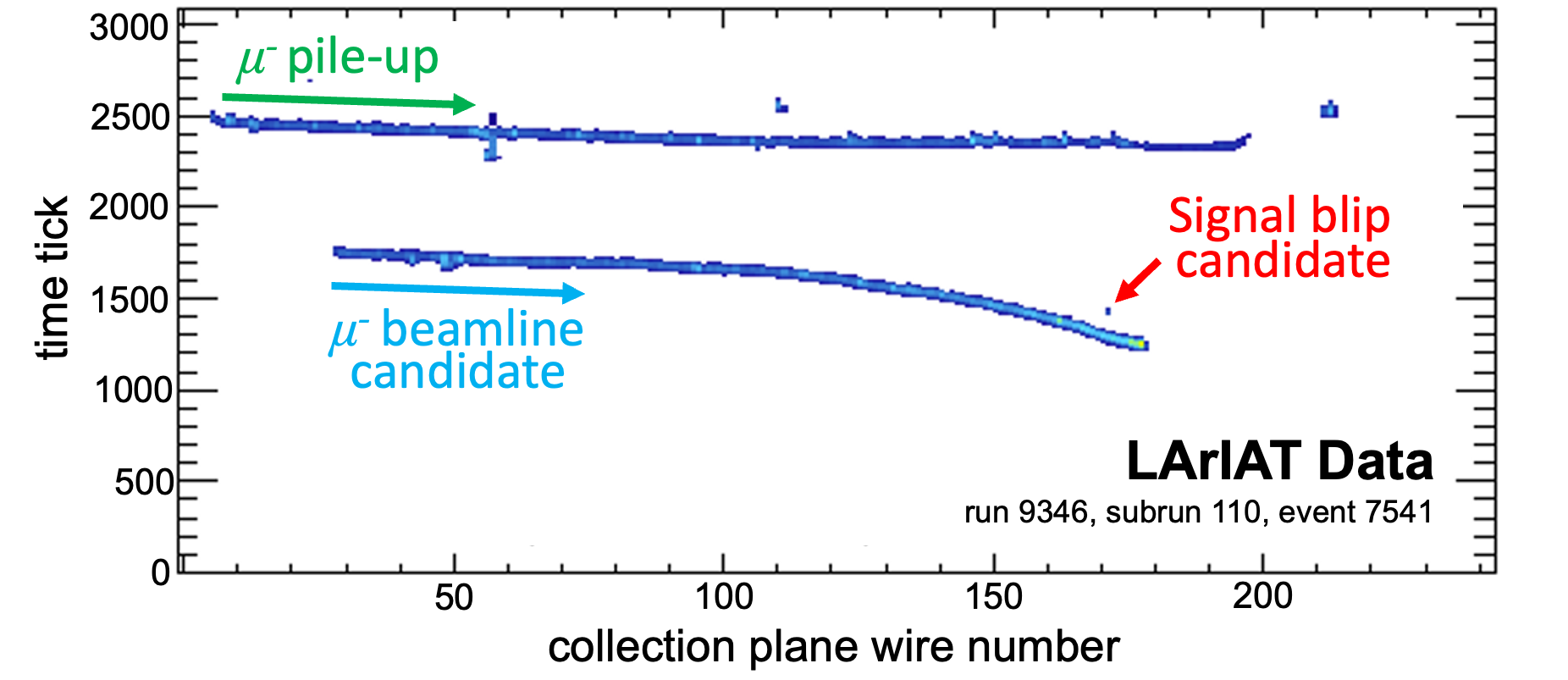}
\caption{A raw LArIAT data event display showing a candidate muon nuclear capture at rest.  MeV-scale blip activity is visible near the signal track's endpoint.  Background tracks and blips further from this endpoint, caused primarily by pile-up beam muons and neutrons, are also visible in the event.  A rainbow color scale is used to indicate ADC signal amplitude per readout time tick.}
\label{fig:EVD}
\end{figure}

A series of selection cuts on beam wire chamber and TPC track information were used to isolate a pure sample of events with a nuclear capture of a stopped $\pi^-$ or $\mu^-$ within the active TPC.  
To illustrate applied cuts, a selected data event display is shown in Figure~\ref{fig:EVD}.  
First, the entry point of an event's beam particle into the TPC, as projected by the MWPCs, was required to match the upstream starting point of a reconstructed TPC track within a 2~cm radius (referred to as a MWPC-TPC match).  
Next, to study only the subset of these signal beam particles capable of ranging out inside the TPC, only beam particles with momentum $<$415~MeV/c were kept.  
In selected events, the signal TPC track must be $>$35~cm long, and while it must start $<$5~cm from the upstream TPC face, it cannot be throughgoing, \emph{i.e.}, also having an endpoint $<$2~cm from the TPC's back face.  
To keep stopping signal tracks with an observable Bragg peak, the median $\mathrm{d}E/\mathrm{d}x$ of all hits within the last 6~cm of this track was required to be $>$3~MeV/cm.  

\begin{table}[tb!]
\begin{tabular}{l|c|c|c}
    \multirow{2}{*}{Cut}  & \multicolumn{3}{|c}{MC} \\
     & All & $\mu$CAR & $\pi$CAR \\
    \hline
    MWPC-TPC Match  &148834& 3231 & 5769 \\
    \hline
    Beam Momentum & 35654& 3231 & 5562 \\
    \hline
    Signal Track Selection & 35031& 3195 & 5480 \\
    \hline
    Bragg Peak& 10283& 1912& 3619 \\
    \hline
    Total Track Number & 9457& 1793 & 3404\\
    \hline
    Particle Range~~~~~($\mu$) & 2132 & 1686 & 3 \\
    Requirement~~~~~~~\,($\pi$) & 3931 & 61 & 2984 \\
    \hline
    \hline
    \multirow{1}{*}{Selected~~~~~~~~~~~~~~($\mu$)}  & 0.014 & 0.52 & $<$0.01 \\
     Fraction~~~~~~~~~~~~~~($\pi$)& 0.026 & 0.02 & 0.52 \\
    \hline
    \multirow{2}{*}{Purity}~~~~~~~~~~~~~~~~\,($\mu$) & - & 0.79 & - \\
     ~~~~~~~~~~~~~~~~~~~~~~~~~($\pi$) & - & - & 0.76 \\
    \hline
    \end{tabular}
\caption{Summary of signal selection cuts and associated data reduction and signal purity expectations.  Event counts reduce moving down the table as cuts are successively added.}
\label{tab:selection}
\end{table}

Coincident activity from pile-up beam muons was reduced by requiring four or fewer total tracks, and all non-signal throughgoing tracks were required to be $>$8~cm from the signal track.  
To categorize remaining events into $\pi$CAR and $\mu$CAR samples, we compared their reconstructed beam momentum, $p$ (in MeV/c), and TPC signal track range, $L$ (in cm), with boundaries optimized based on LArIAT beam particle MC simulations described in the following section.  
As illustrated in the accompanying supplementary materials, these categories were determined by evaluating $L$ in relation to the functional form predicting track length based on beam particle momentum, $Ap-B$. Events satisfying $L>0.43p-79.5$ were categorized as $\mu$CAR candidates, while tracks with length ranging between this boundary and 
$L>0.41p-86.5$ were categorized as $\pi$CAR candidates.
These requirements result in the selection of 87 and 209 $\mu$CAR and $\pi$CAR signal events, respectively.  

LArIAT beamline simulation tools, previously described in Ref.~\cite{LArIAT:2021yix}, were used to generate a MC beam dataset roughly five times larger than the analyzed dataset.  
Generation of tertiary beam particles and their propagation downstream to the final wire chamber were simulated using \texttt{G4Beamline}~\cite{4440461}.  
Particle transport and detector response were simulated using LArSoft~\cite{larsoft} v08\_38\_01, which employs Geant4 v4\_10\_3\_p03e~\cite{g4}.  
Beam pile-up muons were also simulated using \texttt{LArSoft}, with counts and trajectories dictated by analysis of real beam data.  
MC-predicted efficiencies for selecting $\mu$CAR and $\pi$CAR events with the analysis cuts above, shown in Table~\ref{tab:selection}, are 52\%, while predicted signal purity is 79\% and 76\%, respectively.  
A large majority of backgrounds in the $\mu$CAR sample are muon decays at rest, while backgrounds for the $\pi$CAR sample are evenly split between $\mu$CAR-terminating tracks, inelastically scattering pions, and pions absorbed in flight.  

With a purified selected sample of nuclear capture events in hand, we turn to studying the reconstructed blips produced by the captures' final-state products.  
To remove activity from track $\delta$ rays and bremsstrahlung radiation, we exclude blips within 2~cm of any reconstructed track, as well as all blips with \Eb$>$3~MeV.  
To avoid complexities in reconstructing activity at a particle's stopping point, we also cut blips appearing within 3~cm of the signal track's endpoint.  
As summarized in Table~\ref{tab:counts}, this selection yields an average of 5.23 and 6.42 blips per event in the selected $\mu$CAR and $\pi$CAR datasets.  

\begin{table}[tb!]
\begin{tabular}{l|c|cc|c}
\multirow{2}{*}{Dataset} & Throughgoing & \multicolumn{2}{|c|}{Signal}  & \multirow{2}{*}{MC} \\ 
            & Background & Raw & Subtracted & \\ \hline
\multirow{2}{*}{$\mu$CAR}  & & 5.23 & 0.93 & 1.34 \\
   & \multirow{1}{*}{4.30} & \textbf{(1.69)} & \textbf{(0.74)} & \textbf{(1.22)} \\ \cline{1-1}\cline{3-5}
\multirow{2}{*}{$\pi$CAR}  &  \textbf{(0.95)} & 6.42 & 2.12 & 2.93 \\
  & & \textbf{(2.81)} & \textbf{(1.86)} & \textbf{(2.34)} \\
    \end{tabular}
\caption{Blip counts per event for background data and for signal data and MC. Bolded parenthesized values indicate counts for the volume $<$25~cm from the signal track endpoint.}
\label{tab:counts}
\end{table}

\begin{figure}[t!]
\includegraphics[width=\columnwidth]{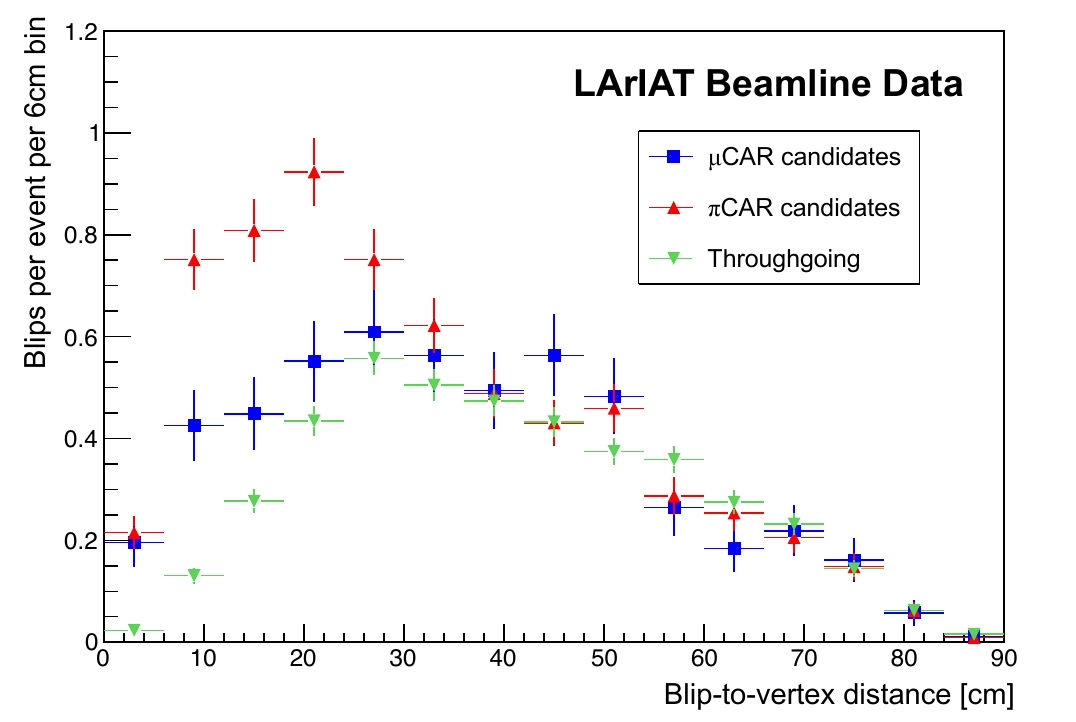}
\caption{Distance between reconstructed blips and their associated signal track endpoints for muon (blue squares) and pion (red triangles) nuclear capture-at-rest datasets.  This metric is also provided for blips in the throughgoing track dataset, with distances given with respect to a randomly-chosen signal track endpoint as described in the text.  Error bars represent statistical uncertainties. \label{fig:blip_data}}
\end{figure}

Blips in selected events are generated by a combination of true stopped beam particle nuclear captures (signal), beam particles undergoing other nuclear interactions, pile-up beam muons, incorrect matches between unrelated features on induction and collection planes, ambient noise and radiogenic signatures, and pile-up neutrons.  
Blip contributions from the first four categories can be estimated using the LArIAT beam MC simulations.  
To quantify expected ambient noise and radiogenic blip populations, identical blip selection criteria were applied to the off-beam pedestal dataset, yielding a sub-dominant 0.36 blips per event.  
Since pile-up beam neutrons were not simulated, their contributions were estimated using a set of 513 events containing throughgoing signal tracks with measured beam momentum between 440 and 500 MeV/c.  
By subtracting blip attributes of the throughgoing dataset from signal $\mu$CAR and $\pi$CAR datasets, we remove all contributions from sources unrelated to the beam particle's terminating process.  
The average blip count per event in the throughgoing event dataset is also given in Table~\ref{tab:counts}.  
A check of blip counts at $z< 35$ cm, a TPC region hosting almost no signal track endpoints, reveals identical results between throughgoing and combined signal datasets -- averaging roughly 1.57 blips per event within $\pm$5.5\% statistical uncertainty -- indicating that the background-subtraction method works as expected.  
A similar $z<$35~cm requirement applied to the LArIAT MC dataset yields a blip count of only 0.21~per event.
This demonstrates that most background blips in this analysis are produced by the un-simulated population of pile-up beam neutrons,  which is correctly accounted for with the data-driven method described above.  

Figure~\ref{fig:blip_data} shows the distance from each blip to its corresponding track's endpoint for signal and throughgoing events.  
To maintain a similar blip selection efficiency in throughgoing events, which do not contain a track endpoint in the TPC's bulk, this variable is defined for thoroughgoing events by taking the distance from each blip in the event to a signal track endpoint vertex of a randomly chosen signal $\mu$CAR or $\pi$CAR event, similarly to Ref.~\cite{argo_mev}.  As expected, blip counts at long distances are comparable between datasets.  
However, at short distances counts deviate well beyond the dominant statistical errors of the two datasets.  
This difference is highlighted in Table~\ref{tab:counts}, which quotes average blip counts for the region within 25~cm of a signal track endpoint: after background subtraction, an average blip count of 0.74~$\pm$~0.19 and 1.86~$\pm$~0.17 per event is observed near muon and pion capture points, respectively.  
A $\chi^2$ comparison between throughgoing and $\mu$CAR ($\pi$CAR) samples in Figure~\ref{fig:blip_data} yields 47.7 (251.5) for 15 degrees of freedom, a 4.2$\sigma$ ($\gg5\sigma$) confidence level (CL) statistical incompatibility.
A comparison between $\mu$CAR and $\pi$CAR blip samples yields a $\chi^2$ per degree of freedom of 40.86/15, indicating 3.6$\sigma$ incompatibility in these two samples.  
Thus, we have provided the first observation of the products of stopped pion and muon nuclear capture on argon, and we have shown that nuclear capture products can be used to differentiate muon and pion samples in LArTPC data.  

\begin{figure}
\centering
\includegraphics[width=\columnwidth]{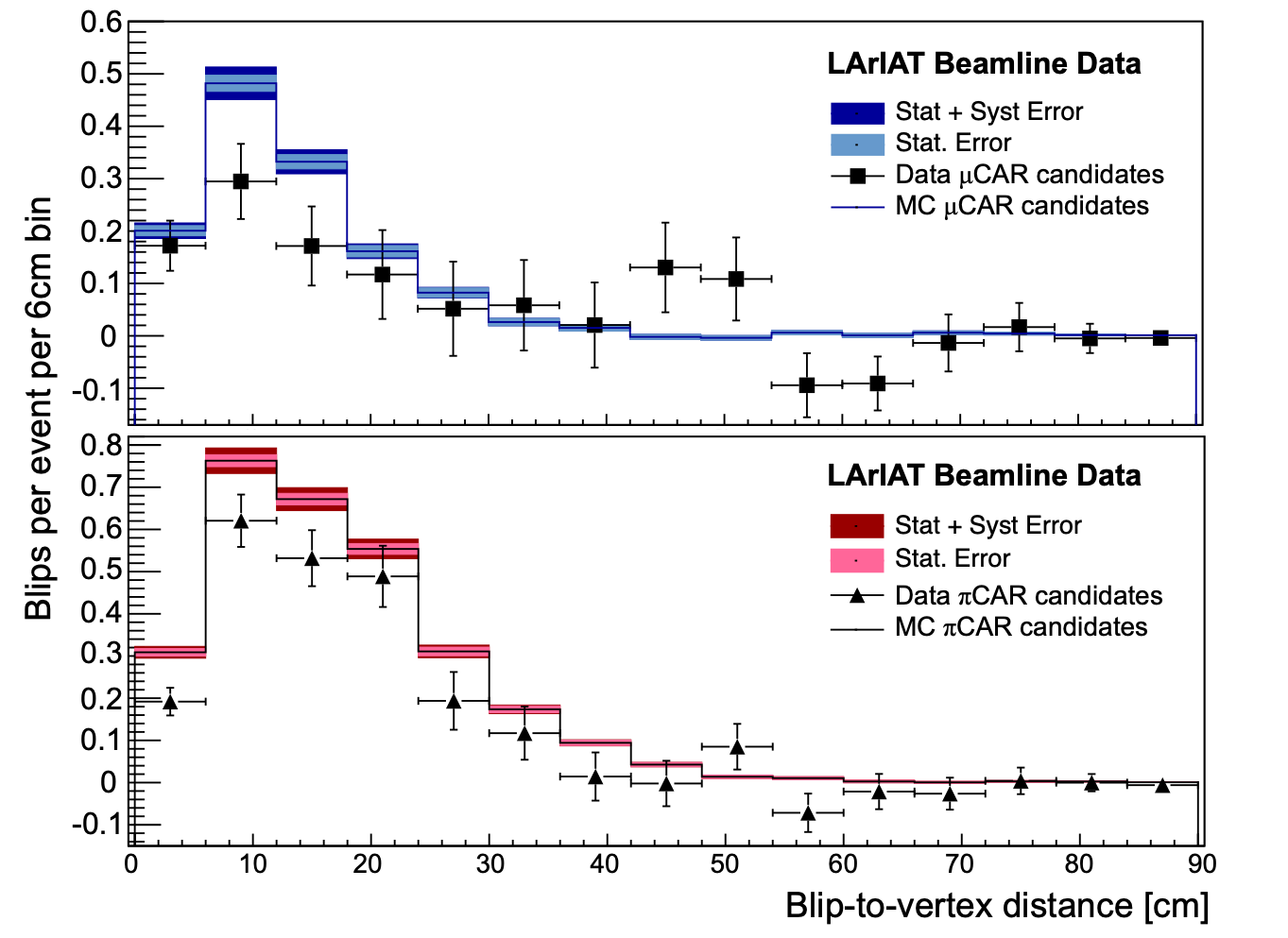}
\caption{Distance between reconstructed blips and signal track endpoints for background-subtracted muon (top) and pion (bottom) nuclear capture-at-rest datasets for data (data points) and Monte Carlo (solid lines).  For data, vertical (horizontal) bars represent statistical errors (bin widths), while for Monte Carlo, statistical and systematic errors are indicated.}
\label{fig:blip_data2}
\end{figure}

A comparison of blip attributes between data and MC for $\mu$CAR or $\pi$CAR signals is provided in Table~\ref{tab:counts} and Figure~\ref{fig:blip_data2}.  
Other blip attributes of interest for data and MC, such as individual blips' \Eb and ($x$,$y$,$z$) positions and events' summed blip count and \Eb, are pictured in supplemental materials accompanying this letter.  
The MeV-scale content present in LArIAT MC events for both muon and pion nuclear captures at rest is substantially larger than that observed in data.  

A detailed comparison of data and MC $\mu$CAR or $\pi$CAR blip samples was performed using a pulls-approach $\chi^2$ test statistic:
\begin{widetext}
\begin{equation}
\label{eq:chi2}
\chi^2 = \sum_{i}\bigg(\frac{[(S^{MC}_i(1+\eta_S)-T^{MC}_i)(1+\eta_C) - (S^{D}_i-T^{D}_i(1+\eta_B))]^2}{\sigma_{\textrm{stat}}^2}\bigg) + \frac{\eta_B^2}{\sigma_B^2} + \frac{\eta_S^2}{\sigma_S^2} + \frac{\eta_C^2}{\sigma_C^2}. 
\end{equation}
\end{widetext}
In this equation, per-event signal ($S$) and background ($T$) content are considered in each vertex-blip distance bin $i$ for data ($D$) and MC.  The statistical uncertainty $\sigma_{\textrm{stat}}$ of each bin considers signal and background contributions for MC and data.   

Nuisance parameters $\eta$ account for the impact of systematic error sources, with pull terms constraining $\eta$ at a level dictated by the size of the associated uncertainty, $\sigma$.  
Background parameter $\eta_B$ is treated as a normalization uncertainty with $\sigma_B$ of 5.5\%.  
Uncertainties in simulated detector response, including electron lifetime, reconstructed energy scales, and blip detection thresholds are accounted for with $\eta_C$; a quadrature-summed normalization uncertainty of $\sigma_C$=3.0\% was defined by observing changes in blip counts in MC samples with varied detector response parameters.  
A signal normalization uncertainty $\eta_S$ is driven by uncertainties in the beam momentum ($p$) and track range ($L$) variables defined above, which are crucial for classifying $\mu$CAR and $\pi$CAR events.  
This source is dominated by a 6.4~MeV uncertainty in $\mu$/$\pi$ energy loss between wire chamber and LArTPC elements, caused primarily by a plastic scintillator halo paddle traversed by a subset of beam particles.  
A 6.4~MeV energy loss (2.6~cm track length) smearing applied to the $\mu$/$\pi$ classification increases blip counts by 3.2\% for $\mu$CAR and decreases counts 1.3\% for $\pi$CAR.  
Thus, $\sigma_S$ is assigned as 3.2\% (1.3\%) for $\mu$CAR ($\pi$CAR), with $\eta_S$ constrained to float only to positive (negative) values.  
Combining and propagating MC statistical errors and $\sigma_B$, $\sigma_S$, and $\sigma_C$ 1$\sigma$ uncertainty values yields 6.2\% (3.9\%) fractional error in predicted total blip counts of 1.22 (2.34) for $\mu$CAR ($\pi$CAR) MC samples.

Calculating the test statistic in Eq.~\ref{eq:chi2} for the data and MC samples, we obtain a $\chi^2$ of 22.07 and 21.32 with 12 degrees of freedom (15 bins with 3 fit parameters) for $\mu$CAR and $\pi$CAR, respectively.  
These $\chi^2$ correspond to p-values of 0.037 and 0.046 for $\mu$CAR and $\pi$CAR, indicating statistical incompatibility of the data and MC at 2.1$\sigma$ and 2.0$\sigma$ CL.  

Based on this $\chi^2$ comparison, it appears that data-MC discrepancies are unlikely to be explained by mis-modeling of detector, beam, or blip background attributes.  
A truth-level selection of only simulated events containing a nuclear capture at rest results in marginal shifts in MC-predicted per-event blip counts (0.1 per event), indicating that mis-modelling of non-CAR $\mu$/$\pi$ nuclear processes also cannot cause the observed over-prediction.  
We thus conclude that Geant4 modelling of the capture process and subsequent nuclear deexcitation is the likely the cause of the discrepancy.  
This conclusion is supported by a comparison of the prevalence of different $\mu$CAR final state nuclei in MC to those provided in Ref.~\cite{muAr}.  
The MC reports 29\% (9\%) $^{40}$Cl ($^{39}$Cl) in the final state, while Ref.~\cite{muAr} measures 7\% (49\%), suggesting under-prediction of pole-term knock-out of the capturing nucleon~\cite{MEASDAY2001243}.  
Meanwhile, the MC reports 36\% of final-state nuclei having A$<$38, while this population is entirely absent in Ref.~\cite{muAr}, indicating incorrect modelling of nuclear evaporation and other deexcitation processes.  

In summary, we have used the LArIAT LArTPC to perform first measurements of the final-state products of $\pi^-$ and $\mu^-$ nuclear captures at rest on argon.  
In 209 $\pi$CAR and 87 $\mu$CAR candidate events, final-state blip activity is observed above background at high ($>$4$\sigma$) confidence level, and a difference in blip content between $\pi$CAR and $\mu$CAR is observed with similarly high (3.6$\sigma$) confidence.  
This represents the first demonstration of the utility of blip information for performing particle-type and charge-sign discrimination for pions and muons in LArTPCs.  
Geant4-based MC simulations of $\mu$CAR and $\pi$CAR in LArIAT cannot accurately reproduce the observed blip activity.  
Thus, improved modelling of these medium-energy nuclear processes is essential in order to fully realize this new and promising LArTPC particle discrimination capability.




\acknowledgments
This document was prepared by the LArIAT collaboration using the resources of Fermilab, a U.S. Department of Energy, Office of Science, HEP User Facility. Fermilab is managed by Fermi Research Alliance, LLC (FRA), acting under Contract No. DE-AC02-07CH11359. We extend a special thank you to the coordinators and technicians of the Fermilab Test Beam Facility, without whom this work would not have been possible. This work was directly supported by the National Science Foundation (NSF) through Grant No. PHY-1555090. We also gratefully acknowledge additional support from the NSF; Brazil CNPq Grant No. 233511/2014-8; Coordena\c{c}\~ao de Aperfei\c{c}oamento de Pessoal de N\'ivel Superior - Brazil (CAPES) - Finance Code 001; S\~ao Paulo Research Foundation - FAPESP (BR) grant number 16/22738-0; the Science and Technology Facilities Council (STFC), part of the United Kingdom Research and Innovation; The Royal Society (United Kingdom); the Polish National Science Centre Grant No. Dec-2013/09/N/ST2/02793; the JSPS grant-in-aid (Grant No. 25105008), Japan; and the Science and Technology Facilities Council (STFC) grant number ST/W003945/1. Miguel Hernandez-Morquecho was supported by the Visiting Scholars Award Program of the Universities Research Association and the Starr-Fieldhouse Research Fellowship at the Illinois Institute of Technology. 
\bibliography{refs}{}

\appendix
\begin{widetext}
\newpage
\textbf{\Large{Supplementary material for ``Measurements of Pion and Muon Nuclear Capture at Rest on Argon in the LArIAT Test Beam Experiment"}}

\section{Muon and Pion Beam Particle Range Requirements} 

In this Appendix, we provide further information regarding the discrimination of $\pi$CAR and $\mu$CAR candidates using a combination of beamline-measured momentum and TPC-measured track length information, which was initially discussed on the manuscript's third page.  
An incoming particle momentum variable, $p$, in units of MeV/c, was determined by measuring the bending radius of the tracked beam particle on its path through the beamline instrumentation.   
The track range variable for an identified signal TPC track, $L$, in units of cm, was determined by the 3D clustering and tracking algorithms used in LArIAT's automated reconstruction software.  

To compose segregated $\pi$CAR- and $\mu$CAR-including event samples, we performed an analysis of $p$ and $L$ variables for the LArIAT beam Monte Carlo (MC) sample described in the manuscript.  
Figure~\ref{fig:class} shows the combination of $p$ and $L$ values for beam MC signal particles passing all other event selection cuts described in the manuscript.  
Three separate populations are visible.  
Two populations exhibit a clear, linear relationship between $p$ and $L$, one with a lower associated momenta (presumably stopping lower-mass muons) and one with a slightly higher associated momenta (presumably stopping higher-mass pions).  
These two populations appear well-separated from one another, indicating the ability to apply high-efficiency, high-purity cuts in this phase space.  
A third population of events with comparatively shorter track length exhibits little correlation between $p$ and $L$ variables, suggesting that this population contains particles that interact prior to stopping in the TPC.  
A truth-level study of these events, described briefly in the manuscript, shows that true identities and terminating processes for particles in these three populations accord well with the descriptions given above.  

\begin{figure} [ht]
\centering
\includegraphics[width=0.65\textwidth]{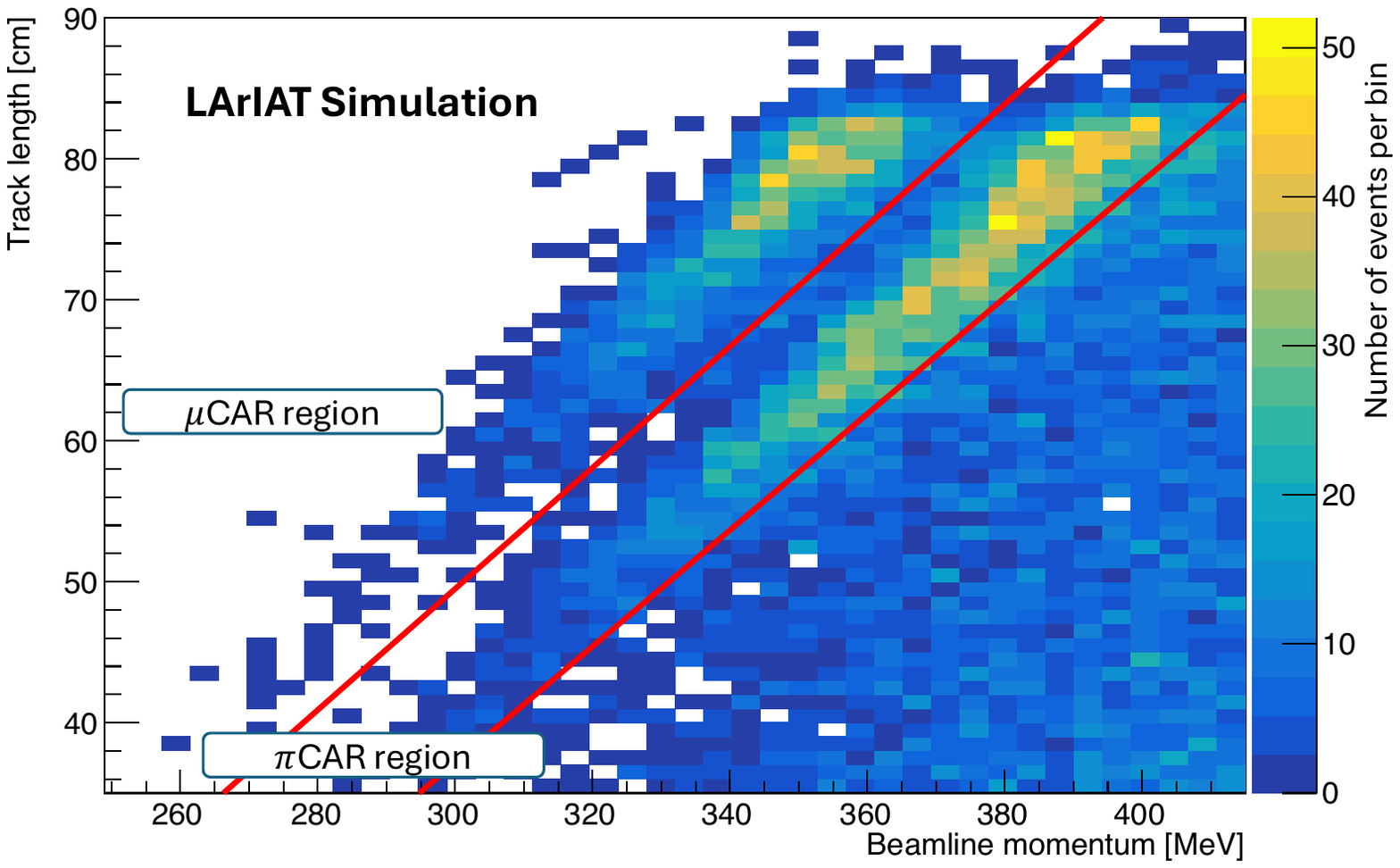}
\caption{To select between $\mu$CAR and $\pi$CAR, we use MC to evaluate the relationship between the reconstructed track length of the particle in the TPC, $L$, and its associated momentum measured in the beamline, $p$. There are two signal regions: one on the left where $\mu$CAR is dominant with a purity of 79\%, and one in the middle where $\pi$CAR is dominant with a purity of 76\%.}
\label{fig:class}
\end{figure}

Final chosen $\pi$CAR and $\mu$CAR discrimination boundaries in $L$-$p$ space, defined in the manuscript, are also depicted in Figure~\ref{fig:class}.  
The MC-reported efficiency and purity of the resultant $\pi$CAR and $\mu$CAR sample are also reported in the main manuscript.  

\newpage

\section{Other Attributes of Selected Signal Blips}

In this appendix, we provide further description of the properties of blips observed in signal $\pi$CAR and $\mu$CAR events.  

\begin{itemize}
    \item \textbf{Truth-level blip content}: Blips' individual energies are depicted in Figure~\ref{fig:blip_energyr1} and blips' distances to their associated signal track endpoint are shown in Figure~\ref{fig:blip_tovertexr1}.  
    In these distributions, points represent the data after performing the background subtraction procedure described in the manuscript; negative values thus indicate upward statistical fluctuations present in the subtracted throughgoing data sample.  
    Stacked histograms in these figures represent the MC-predicted contributions of various blip-generating interaction processes, which include the signal CAR process, muon decays at rest (DAR), other inelastic pion interactions in flight, muon bremsstrahlung, incorrect matches between planes, and other sub-dominant processes.    While the same background-subtraction procedure was applied the beam MC sample in the analysis described in the manuscript (and in Figure~\ref{fig:blip_data2}), Figures~\ref{fig:blip_energyr1} and ~\ref{fig:blip_tovertexr1} do not include background subtraction for MC.  
    While this choice modestly reduces the value of a direct data-MC comparison in these figures, it enables better clarity in understanding the primary origin of signal blips within 
    simulated $\pi$CAR- and $\mu$CAR-including events.  
    
    \item \textbf{Blip spatial distribution}: Figure \ref{fig:blip_x} depicts the distributions of reconstructed blips’ $x$, $y$, and $z$ coordinates for data and MC nuclear capture-at-rest datasets.  The background-subtraction procedure has been applied to both MC and data in all three figures.  To allow easier comparison of differential trends between data and MC despite the overall blip count offset between the two reported in the manuscript, we have normalized MC distributions to match their data counterparts.  

    \item \textbf{Total blip energy and multiplicity per event}: The total blip energy and multiplicity per event is evaluated for all signal blips less than 25 cm in distance from their signal track endpoint.  Figure~\ref{fig:totalblip_energy} shows total summed blip energy and multiplicities for $\pi$CAR and $\mu$CAR samples.  
    We note that no background subtraction procedure has applied to these variables, so summed variables include background contributions, specifically from dominant pile-up neutrons.  Table~\ref{tab:counts} provides a sense of the relative contributions of signal and background for these summed blip samples.  Given the substantial difference in data- and MC-reported background contributions due to the lack of pile-up neutron simulation, we choose not to depict MC-predicted summed variables in these figures.  
\end{itemize}

\begin{figure}[ht]
\centering
\includegraphics[width=0.49\textwidth]{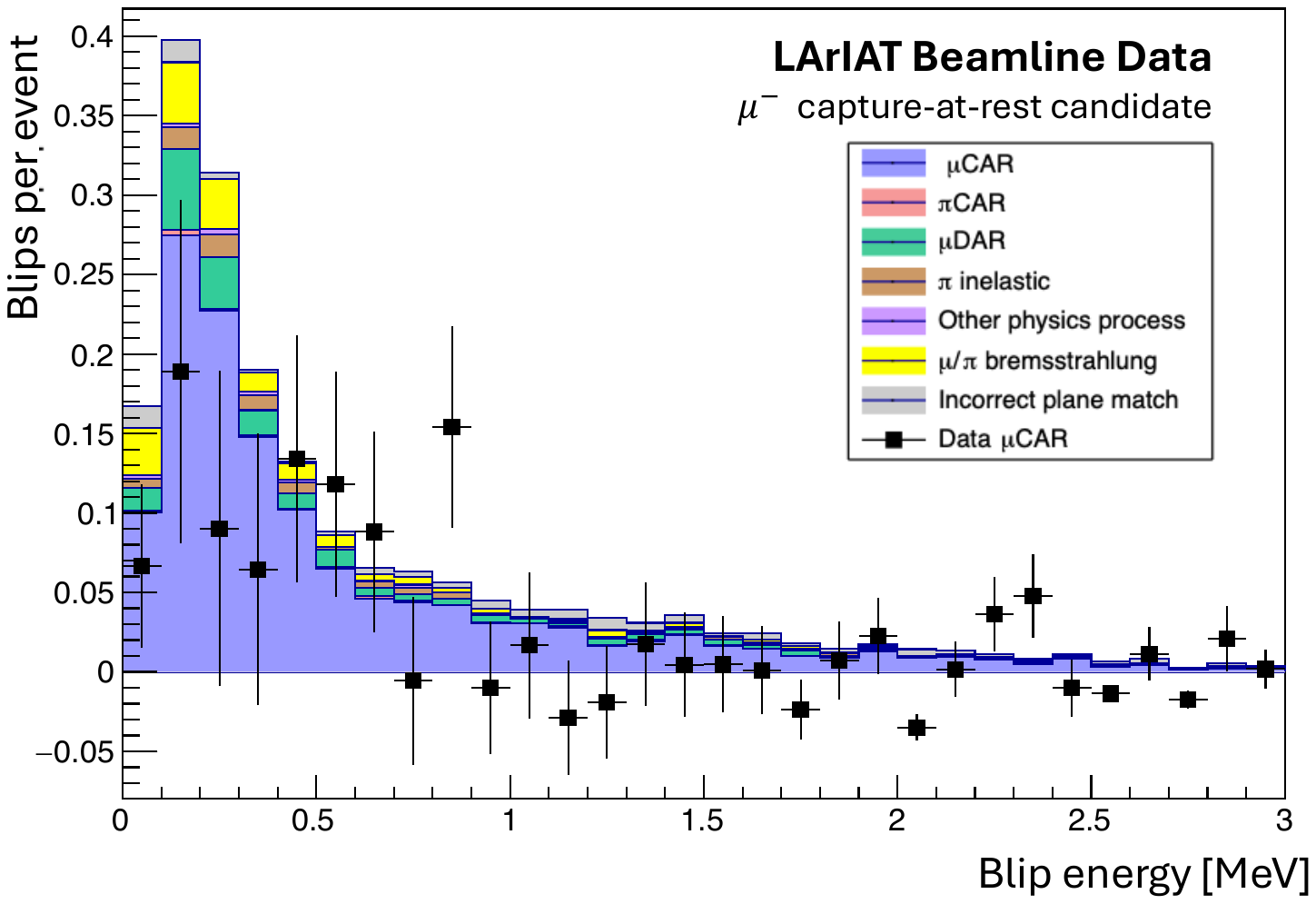}
\includegraphics[width=0.49\textwidth]{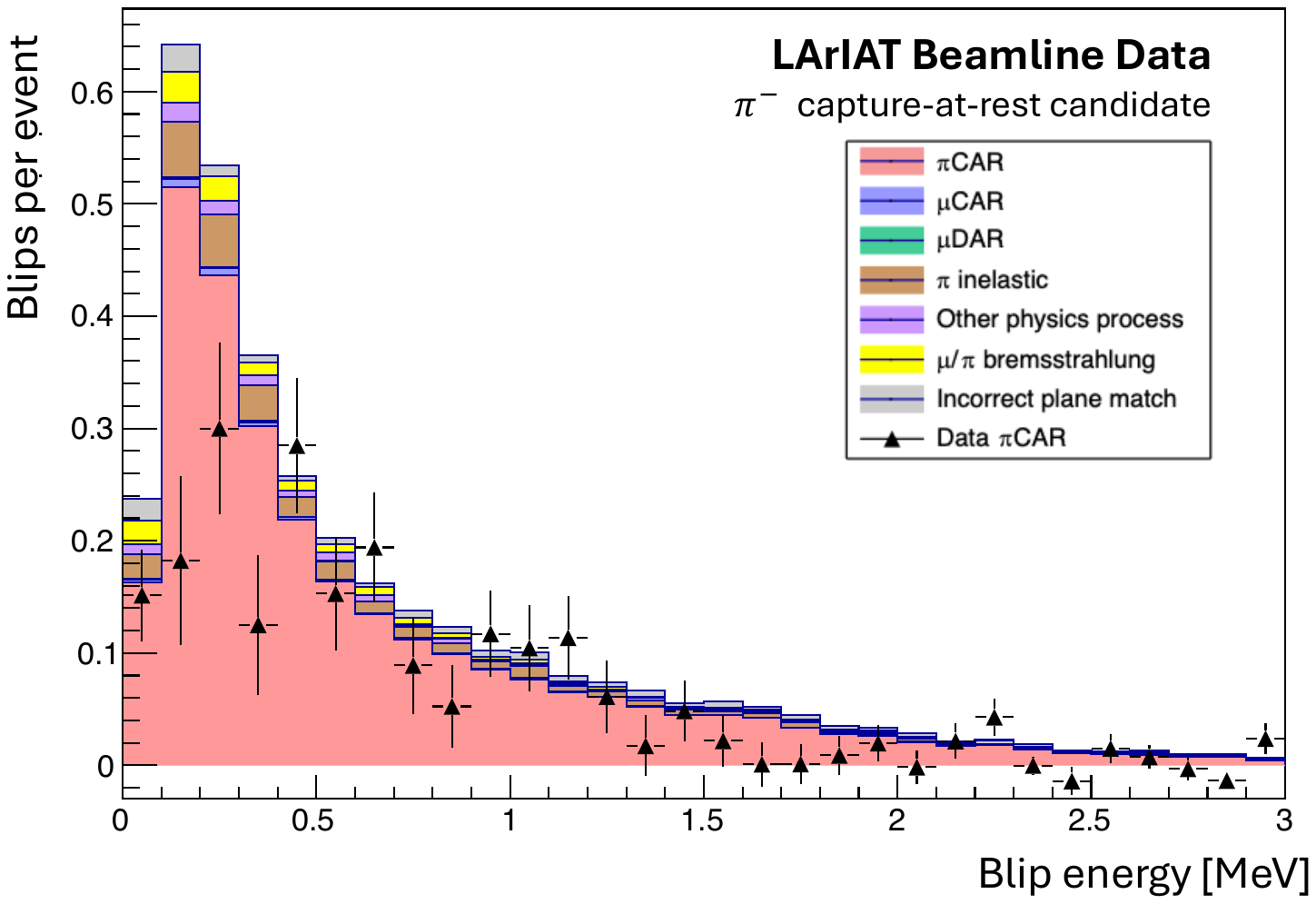}
\caption{The individual energy of blips in MC (stacked histograms) and measured (black) datasets for muon (left) and pion (right) nuclear captures at rest. The MC dataset is labeled according to the true underlying physics responsible for generating each blip.  Error bars depicted in the figures correspond to statistical uncertainties.}
\label{fig:blip_energyr1}
\end{figure}

\begin{figure}
\centering
\includegraphics[width=0.49\textwidth]{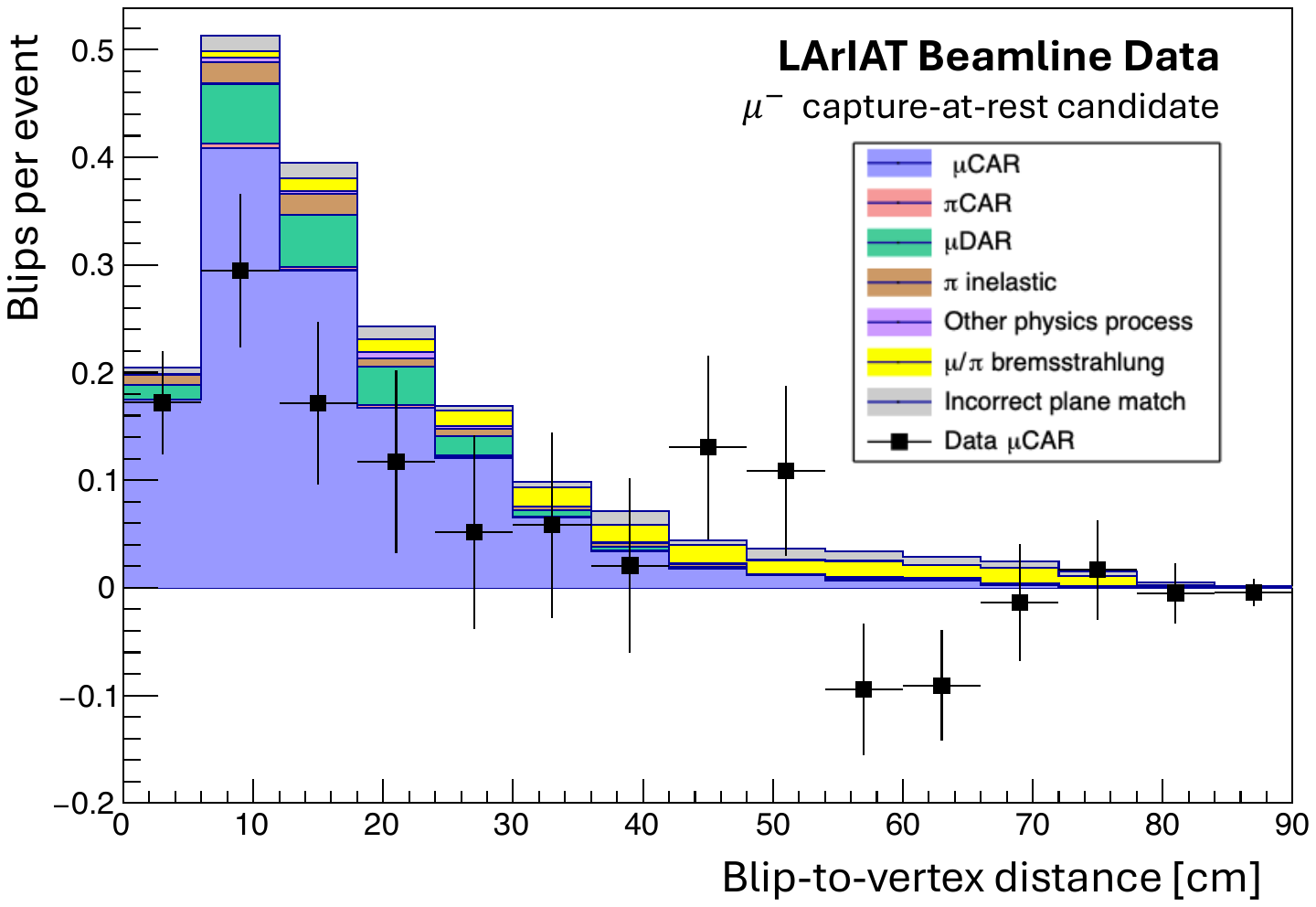}
\includegraphics[width=0.49\textwidth]{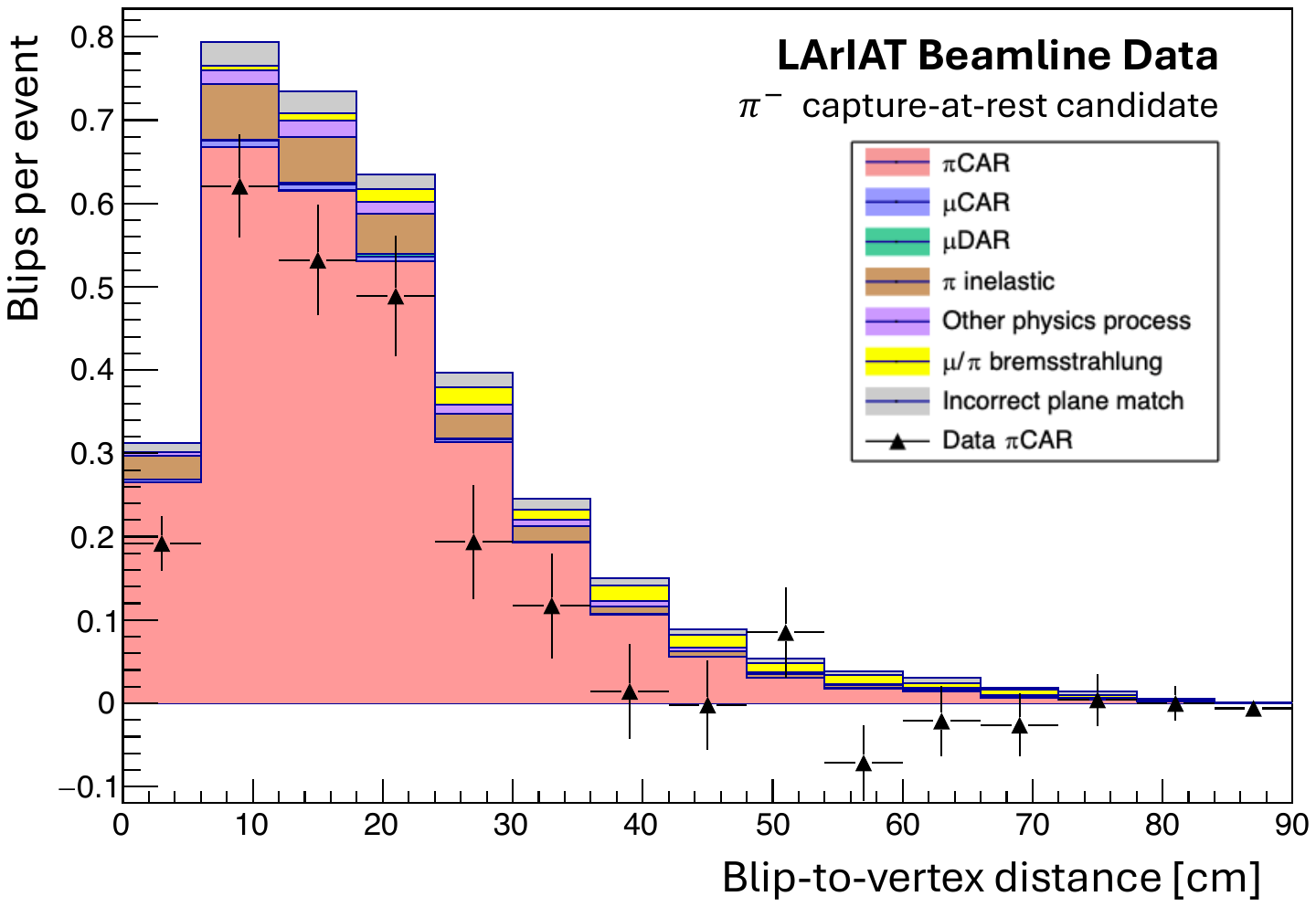}
\caption{Distance between reconstructed blips and signal track endpoints for blips in MC and measured (black) datasets for muon (left) and pion (right) nuclear capture at rest. The MC dataset is labeled according to the true underlying physics responsible for generating each blip.  Error bars depicted in the figures correspond to statistical uncertainties.}
\label{fig:blip_tovertexr1}
\end{figure}

\begin{figure} 
\centering 
\includegraphics[width=0.48\textwidth]{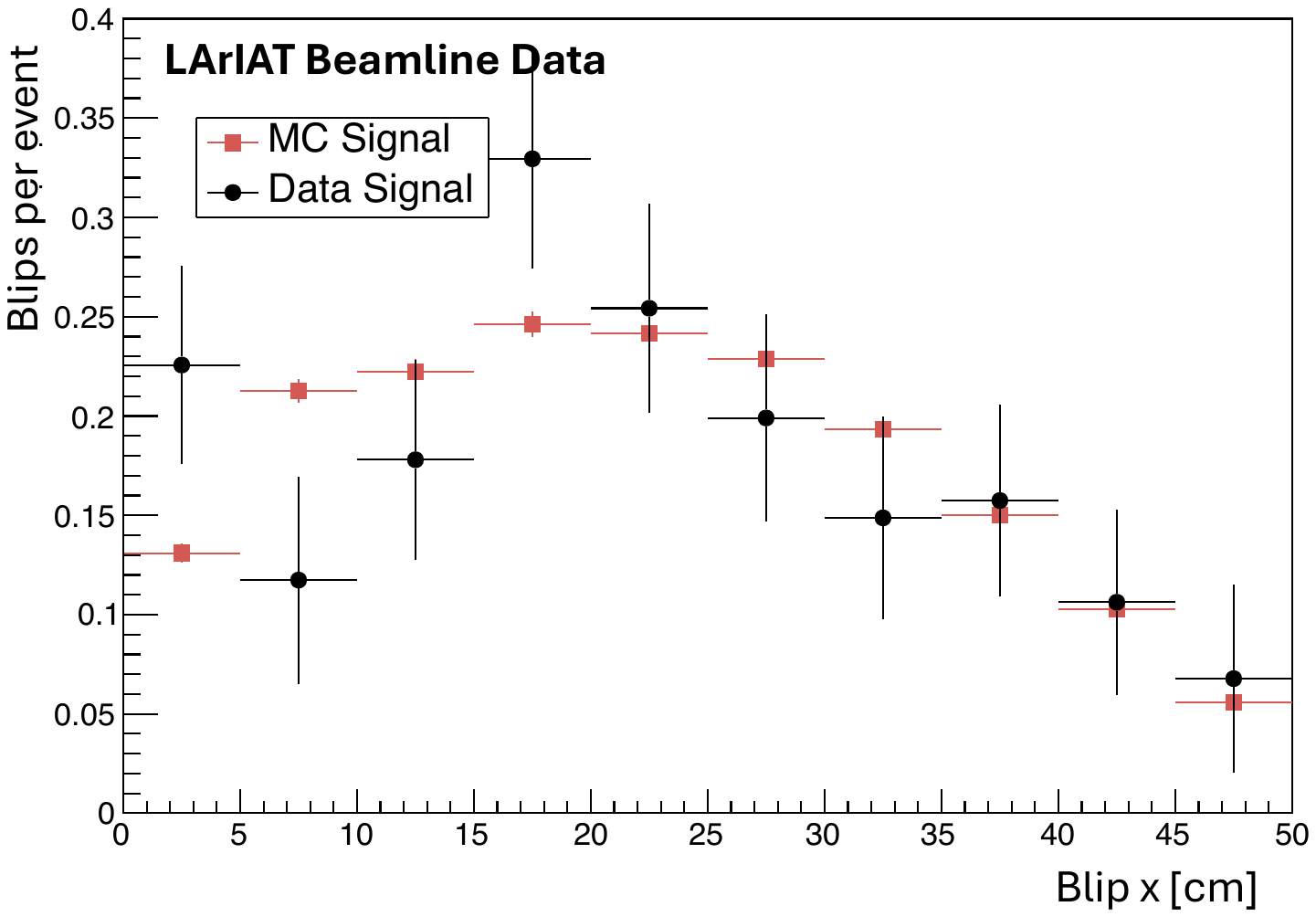}
\includegraphics[width=0.48\textwidth]{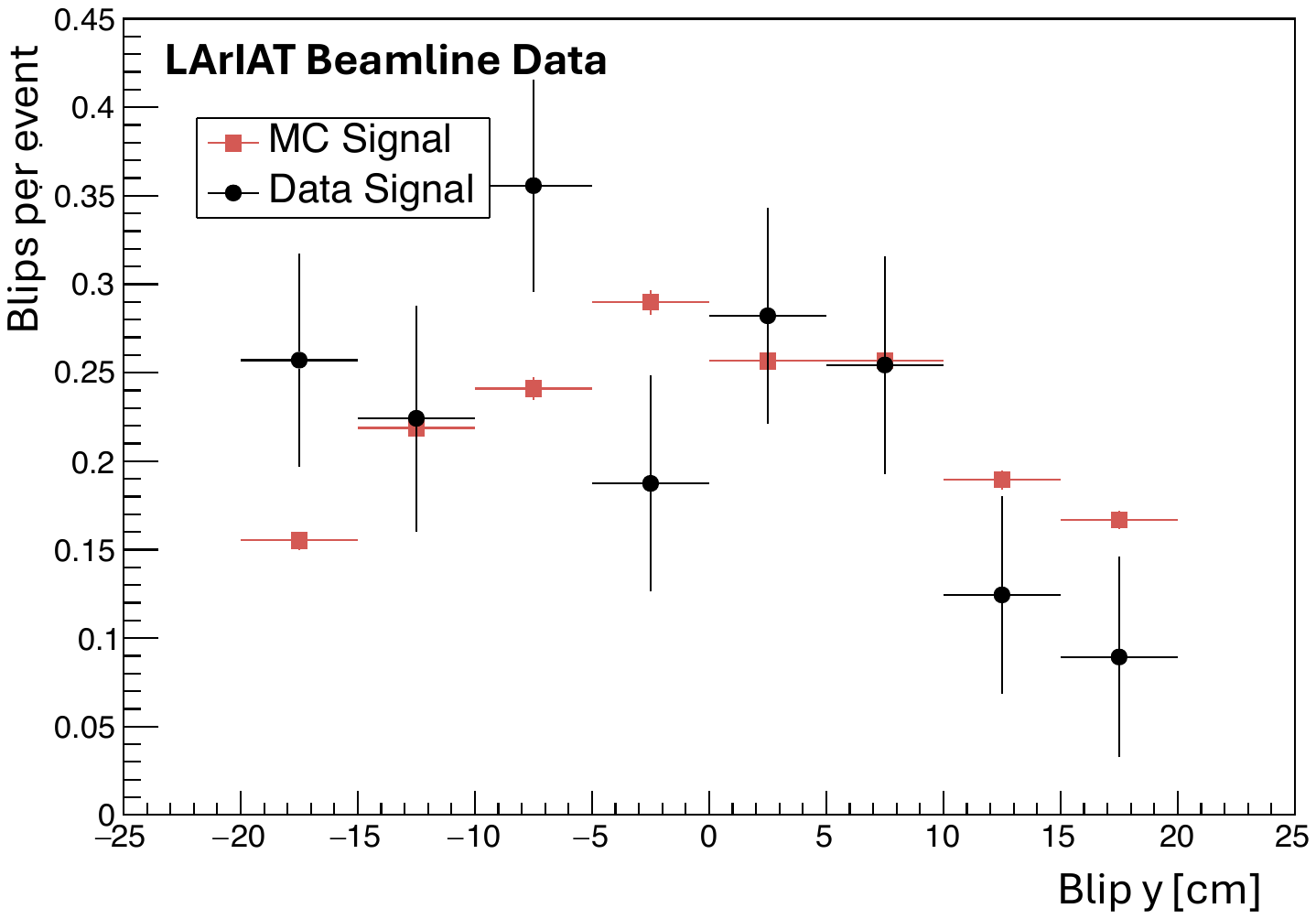}
\includegraphics[width=0.48\textwidth]{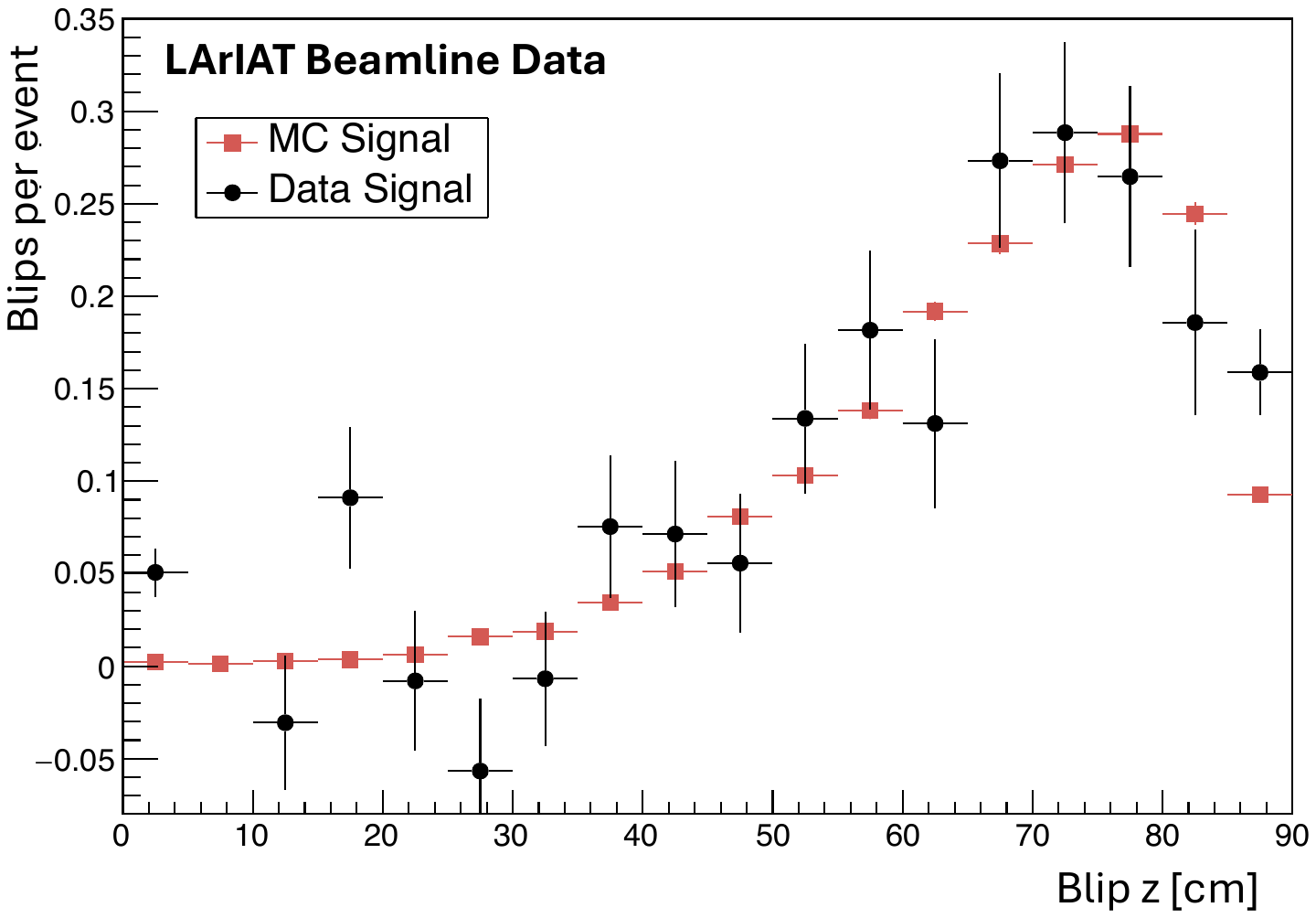}
\caption{Reconstructed $x$ (top left), $y$ (top right), and $z$ (bottom) blip position per event for data (black) and MC (red) nuclear capture-at-rest datasets.  For all panels, $\pi$CAR- and $\mu$CAR samples are summed.  Error bars represent statistical uncertainties.  To enable comparison of differential trends, the MC was normalized with respect to the data.}
\label{fig:blip_x}
\end{figure}

    \begin{figure} 
\centering
\includegraphics[width=0.49\textwidth]{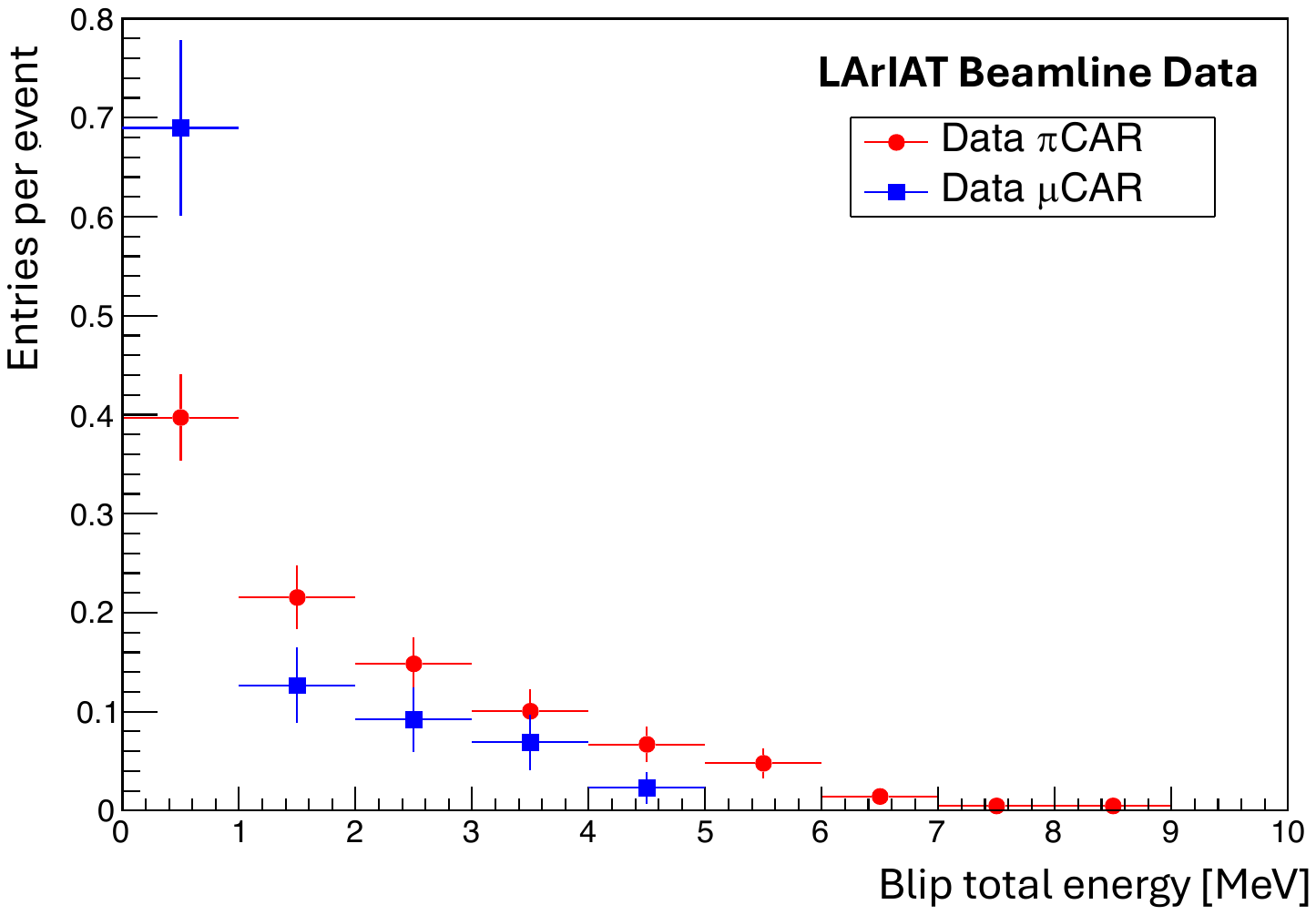}
\includegraphics[width=0.49\textwidth]{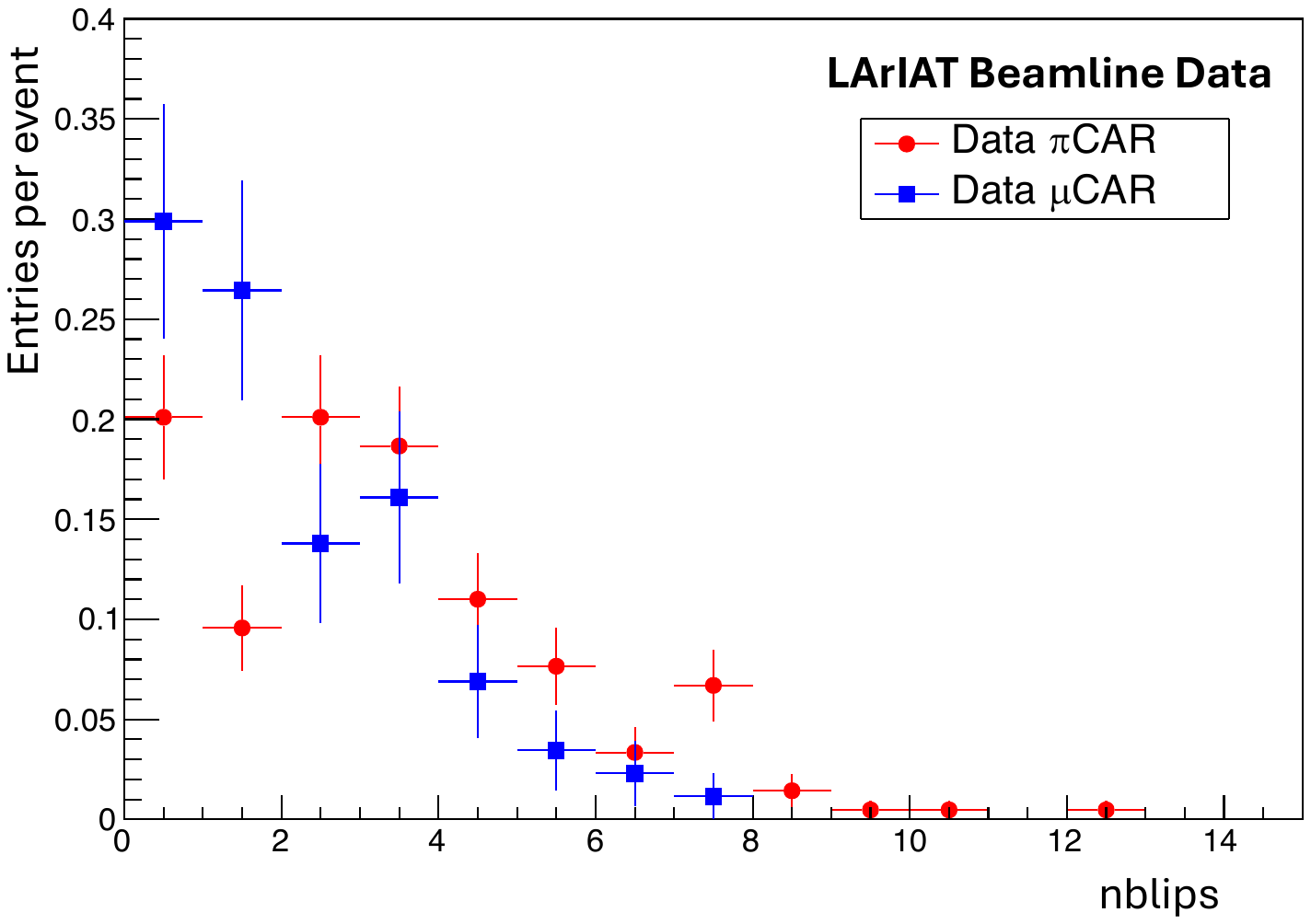}
\caption{The total summed energy (left) and multiplicity (right) of signal blips within 25 cm of the candidate track’s endpoint, plotted separately for muon (blue) and pion (red) nuclear capture-at-rest candidates in the datasets. The error bars represent statistical uncertainties.}
\label{fig:totalblip_energy}
\end{figure}

\end{widetext}

\end{document}